# New frontier of laser particle acceleration: driving protons to 93 MeV by radiation pressure

I Jong Kim[1,2], Ki Hong Pae[1], Chul Min Kim[1,2], Il Woo Choi[1,2], Chang-Lyoul Lee[2], Hyung Taek Kim[1,2], Himanshu Singhal[1], Jae Hee Sung[1,2], Seong Ku Lee[1,2], Hwang Woon Lee[1], Peter V. Nickles[3], Tae Moon Jeong[1,2] and Chang Hee Nam[1,4*]

[1]Center for Relativistic Laser Science, Institute for Basic Science (IBS), Gwangju 500-712, Korea

[2]Advanced Photonics Research Institute, GIST, Gwangju 500-712, Korea

[3]Max Born Institute, Berlin, Germany

[4]Department of Physics and Photon Science, GIST, Gwangju 500-712, Korea

*chnam@gist.ac.kr

**The radiation pressure acceleration (RPA) of charged particles has been considered a challenging task in laser particle acceleration[1-3]. Laser-driven proton/ion acceleration has attracted considerable interests due to its underlying physics and potential for applications such as high-energy density physics, ultrafast radiography, and cancer therapy[4-11]. Among critical issues to overcome the biggest challenge is to produce energetic protons using an efficient acceleration mechanism[12]. The proton acceleration by radiation pressure is considerably more efficient than the conventional target normal sheath acceleration driven by expanding hot electrons. Here we report the generation of 93-MeV proton beams achieved by applying 30-fs circularly polarized laser pulses with an intensity of $6.1 \times 10^{20}$ W/cm$^2$ to ultrathin targets. The radiation pressure acceleration was confirmed from the obtained optimal target thickness, quadratic energy scaling, polarization dependence, and 3D-PIC simulations. We expect this fast energy scaling to facilitate the realization of laser-driven proton/ion sources delivering stable and short particle beams for practical applications.**

Laser particle acceleration has been explored to realize compact ultrashort proton accelerators. The acceleration of protons and ions using lasers has the advantages of yielding an extremely high electric field of several MV/μm and ultrashort durations of picoseconds or less for novel scientific and engineering applications. In experiments, the target normal sheath acceleration (TNSA) scheme has been extensively investigated due to its robustness to physical conditions[13,14]. Well characterized high-energy protons generated by TNSA have been used for applications such as the radiographic imaging of electrostatic fields in plasma and irradiation of tumor cells[15,16]. However, the slow scaling of the proton energy $E_p$ to the laser intensity $I$, given by $E_p \propto I^{0.5\text{-}1}$, is, a severe limitation of the TNSA in increasing the proton energy[17,18]. In this regard, the RPA scheme has been preferred due to its fast scaling according to the relation $E_p \propto I^2$ [19].

The experimental implementation of RPA, performed by irradiating ultrathin targets with ultra-intense laser pulses, is a challenging task requiring delicate experiments under extreme conditions. If the target is sufficiently thin, electrons can be pushed forward as a whole by the radiation pressure of the laser light, and the resulting charge separation field can pull protons and ions also as a whole. The collective displacement of the electrons by the radiation pressure of an ultra-intense laser pulse and the consequent collective acceleration of protons are clearly seen in Fig.1(a), obtained from a 3D particle-in-cell (PIC) simulation. More specifically the radiation pressure induces a double layer structure of opposite charges, of which an electrostatic field accelerates the protons, as shown in Fig.1(b). The plasma needs to be thicker than the skin depth of the target for achieving significant reflections. Since the laser momentum transfer to electrons is proportional to laser intensity, the energy of protons, pulled by the electrons, increases as the laser intensity squared. For such acceleration to be effective, the double layer structure should be kept as stable as possible over the laser-pulse duration, for which the radiation pressure and the electrostatic pressure on the electron layer should be balanced. This balancing condition yields the optimum RPA condition, $a = \pi \cdot \sigma$, where $a$ denotes the amplitude of the normalized vector potential of the laser pulse and $\sigma$ the normalized areal density of electrons (See Methods). In addition circularly polarized (CP) laser pulses are preferred over linearly polarized (LP) laser pulses since CP pulses quench the longitudinal electron oscillations, which can destabilize the double layer structure. Consequently, super-intense CP laser pulses with extremely high contrast should be delivered to targets, slightly thicker than the skin depth, for the realization of RPA, which makes the experimental demonstration of RPA challenging.

Here we report the experimental demonstration of RPA by applying ultra-intense CP laser pulses to ultrathin (10 nm) polymer targets. Polymer targets are preferred over diamond-like carbon or metals because of their relative proton-richness and low electron densities. The RPA of protons could be clearly confirmed from the proton energy scaling to laser intensity, optimal target thickness, laser polarization dependence, and 3D-PIC simulations. Our achievement of energetic protons via RPA mechanism indicates that laser energy transfer to particles can be controlled in the relativistic regime and that RPA can form a practical scheme for a laser-based proton accelerator in the 100-MeV range.

The RPA of protons was realized after careful optimization of experimental parameters. High-contrast, ultra-intense CP laser pulses were applied to ultrathin polymer targets. The resulting proton and carbon-ion spectra are shown in Fig. 2(a), obtained from a 15-nm-thick polymer target subjected to a CP laser pulse with an intensity of $6.1 \times 10^{20}$ W/cm$^2$. The maximum proton energy was as high as

93 MeV, while the maximum $C^{6+}$ energy was 174 MeV (see also Supplementary Figure S1). In Fig. 2(b), the proton spectra (the cases yielding the highest proton energy with LP and CP laser pulses) show modulated plateau profiles. The proton spectrum modulation may be attributed to the Rayleigh–Taylor (RT) instability[20,21]. The plateau energy profile enables the generation of a larger amount of energetic protons than that generated using the exponentially decreasing profile of the TNSA process. The blue curve, just below the parabolic trace of protons in Fig. 2(a), was used to estimate the signal level of the background. The maximum proton energy obtained with the CP laser pulse (93 MeV) is considerably higher than that with the LP laser pulse (67 MeV), which is a strong indication of the RPA process; it is confirmed further with other evidences.

We investigated crucial features of RPA by controlling the target thickness and laser intensity. The condition for the optimum target thickness for RPA is satisfied when the radiation pressure on electrons balances the electrostatic pressure between the double layers of electrons and ions. In case of targets thicker than the optimum thickness, the double layer structure collapses when the laser pulse penetrates beyond the optimum thickness, since the electrostatic pressure exceeds the radiation pressure. For thinner targets, the electrons are pushed too far from the protons, and they diffuse spatially before the protons are pulled effectively by the electrons, thereby resulting in a weak Coulomb field. For targets thinner than the skin depth (9 nm for our target), the laser pulse is more transmitted than reflected, which significantly reduces the amount of transferred laser momentum. By varying the target thickness over 10 – 100 nm, the optimum target conditions were experimentally determined, as shown in Fig. 3(a). For a CP laser pulse of intensity $6.1 \times 10^{20}$ W/cm$^2$, the highest proton energy of 93 MeV was obtained from a 15-nm target, which thickness value is close to the theoretical estimation of 16 nm. Our 3D-PIC simulations also exhibited the same optimum thickness. On the other hand, for a LP laser pulse of intensity $7.0 \times 10^{20}$ W/cm$^2$, the highest proton energy of 67 MeV was obtained from a 20 nm target. This thickness also agreed closely with the theoretical estimation of 23 nm for LP pulses. The close consistency between the measured optimum thickness and the predicted value of the RPA model, along with the 3D-PIC simulation results, strongly indicates that the double layer structure caused by charge separation was maintained stably via the balancing of the radiation pressure with the electrostatic pressure, thereby generating energetic protons.

The most representative feature of RPA is the quadratic proton-energy scaling with respect to the laser intensity. At the optimum thicknesses of 15 nm for CP laser pulses and 20 nm for LP laser pulses, we measured the proton energy spectra to obtain the proton scaling characteristics for laser intensities of $1.7 \times 10^{20}$ W/cm$^2$ – $7.0 \times 10^{20}$ W/cm$^2$. For the CP laser case, a clear quadratic increase was observed, as shown in Fig. 3(b). This quadratic scaling confirms that the radiation momentum is transferred to protons through electrons pushed collectively by radiation pressure, as predicted by the

RPA model[2, 19]. In contrast, a linear scaling was observed in the case of LP laser pulses. Importantly, the result obtained with CP laser pulses exhibits a better performance than the LP-laser results for intensities above $5.4 \times 10^{20}$ W/cm$^2$ ($a_0$ = 16, see methods). This result agrees with the prediction of Ref. 10, wherein it is suggested that the maximum proton energy with CP laser pulses is higher than that with LP laser pulses for $a_0 \geq 15$. Unlike LP laser pulses, CP laser pulses suppress the transfer of the laser energy to the thermal energy of electrons, which does not contribute to acceleration. Such a clear difference between the LP and CP cases, not previously reported[22,23], confirms the advantage of CP laser pulses in RPA. Consequently, our results, which exhibit the quadratic intensity scaling of the proton energy and outperformance of the CP case over the LP, strongly support our claim on the experimental demonstration of RPA, opening a realistic pathway for producing highly energetic protons.

We note that the observed spectral shape is not quasi-monoenergetic, contrary to the RPA model predictions and some PIC simulations[2,3]. This discrepancy can be attributed to the nonrealistic conditions assumed in the analytic model and the simulations, such as a super-Gaussian or even flat-top laser profile, higher laser intensities beyond $10^{21}$ W/cm$^2$, and reduced physical dimensions[3, 19, 24]. As a result, a quasi-monoenergetic spectrum could not be achieved experimentally as yet to the level of the simulation results or of practical significance[22, 23]. Unless the abovementioned ideal conditions are fulfilled experimentally, the generation of quasi-monoenergetic protons would remain challenging.

In order to examine the underlying physics of RPA, 3D-PIC simulations were performed for the experimental conditions. For the case of the 15-nm target driven with a CP laser pulse of intensity $6.2 \times 10^{20}$ W/cm$^2$, we calculated the evolution of the proton phase space distribution (Fig. 4(a)). The figure clearly shows that protons were well bunched in the phase space, signifying that most protons were accelerated as a whole and that protons followed a spiral motion around the moving phase center determined by collective electrons, as also observed in other simulations[10, 25]. As RPA is effective only during the existence of the laser pulse, further acceleration of protons occurred due to the charge imbalance after the laser-pulse termination. The final proton energy was thus determined by RPA and the post acceleration. As shown in the blue dots of Fig. 4(b), 86 % of the final energy was contributed by RPA in the CP case. Consequently, these detailed features from PIC simulations evidence RPA as the main acceleration mechanism in our experimental results.

In conclusion, we have experimentally demonstrated the RPA of protons through the interaction of ultrahigh contrast CP laser pulses with ultrathin targets. The quadratic scaling of the proton energy to the laser intensity and the optimal target thickness agreed closely with the predictions of the RPA model. It was shown that the proton acceleration with CP laser pulses was considerably more effective than that with LP laser pulses. By applying CP laser pulses with an intensity of $6.1\times10^{20}$ W/cm$^2$ to a

15-nm polymer target, we produced 93-MeV protons. This result, i.e., the clear realization of RPA and the production of energetic protons, is a milestone in laser particle acceleration, paving the way to the "magic" 100-MeV proton energy threshold, which is a key requirement for proton oncology. With upcoming ultra-intense lasers and the precision characterization of energetic particles, laser proton/ion acceleration would promote applications in high-energy density physics, plasma diagnostics, laser nuclear physics, and hadron oncology.

## Methods

### Laser parameters

Experiments were performed using a petawatt (PW) laser delivering an energy of 27 J with a 30-fs pulse duration. The laser pulse was delivered to a double plasma mirror (DPM) system to achieve a high temporal contrast of $3\times10^{-11}$ at 6 ps before the main pulse. The total reflectivity of the DPM system was 33 %, mainly due to the transmission (82 %) through a half-wave plate and the reflectivity (40 %) of the DPM. Subsequent to the DPM stage, a 8.5-J, an s-polarized laser pulse was focused onto ultrathin polymer targets (F8BT) using an off axis parabolic mirror (f/3, f = 600 mm). The measured focal spot size at FWHM was 3.5 μm × 4.6 μm along the horizontal and vertical directions, respectively, and the calculated energy concentration within the FWHM range was about 30 %. Consequently, the maximum intensity on the target was $7.0\times10^{20}$ W/cm$^2$ ($a_0$ = 18) for the case of linear polarization. The laser intensity is usually characterized by the dimensionless parameter $a_0$ = $0.85\cdot\sqrt{I\lambda^2}$, where I denotes the laser intensity in units of $10^{18}$ W/cm$^2$, and λ the laser wavelength in μm.

The amplitude of the normalized vector potential a is given as a = $a_0$ for LP and a = $a_0/\sqrt{2}$ for CP. The laser intensity was controlled from $1.6 \times 10^{20}$ W/cm$^2$ ($a_0$ = 8.7) to $7.0 \times 10^{20}$ W/cm$^2$ ($a_0$ = 18) by adjusting the laser energy. We also installed a mica quarter-wave plate before the OAP mirror to produce CP laser pulses. The intensity of the CP laser pulses was $6.1\times10^{20}$ W/cm$^2$ ($a_0$ = 17) due to the transmission of the quarter wave-plate (87 %).

### Target and diagnostics

The target, which is made of F8BT, is proton-rich and has an electron density around 200 $n_c$, where $n_c$ denotes the critical density. The target thickness was varied from 10 nm to 100 nm. The characteristics of the polymer F8BT target are described elsewhere[26,27]. The normalized areal density of the electrons of a slab target σ is defined by σ = $(n_e/n_c)\cdot(d/\lambda)$ where $n_e$ denotes the electron number density, d the target thickness, λ the laser wavelength, and $n_c$ the critical density for the wavelength

(as defined previously). The target surface was placed at an incident angle of 9.5° with respect to the target normal to minimize back-reflected laser pulses from the targets. For the detection of proton spectra with high energies and charge-to-mass resolutions, a Thomson parabola (TP) spectrometer was used. Parabolic ion traces were recorded using a microchannel plate (MCP) with a phosphor screen imaged to a 16-bit charge-coupled device (CCD) camera. The solid angle of the TP was $3.5\times10^{-8}$ steradians (sr). The magnetic field distribution, including the fringe field, was measured using a Hall probe for the precise estimation of proton energy[28]. The magnetic field in the central magnet region was 0.40 T. For protons with energy up to 6.2 MeV, the calibration of the MCP-phosphor screen-CCD system was performed by installing slotted CR-39 track detectors in front of the MCP[29]. For higher energy protons, we assumed that the efficiency (counts/particle) was constant. The energy measurement error, estimated by considering the line broadening of parabolic ion traces, caused by the finite size of the collimator and the spatial resolutions of MCP and CCD, was about ± 4.7 MeV for 93-MeV protons and smaller for lower energy protons.

**Simulations**

Three-dimensional particle-in-cell simulations were carried out using the relativistic electromagnetic code ALPS. The ALPS code has been successfully used for simulations of high intensity laser-plasma interactions[30, 31]. A laser pulse was launched normally from a boundary with a Gaussian spatial profile (FWHM = 4 μm) and a $\sin^2$ temporal profile (FWHM = 30 fs) with a wavelength of 800 nm. Initially a thin target is positioned at z = 0. The target was modelled by a cold neutral plasma composed of electrons, fully ionized carbon ions ($C^{6+}$), and protons with electron number density $n_e = 209\ n_c$, where $n_c$ denotes the critical density. In simulations, the longitudinal mesh size was set to $\lambda/800$ and the transverse mesh sizes were set to $\lambda/40$, and the macro-particle per cell was set to 20 for electrons and 10 for each ion species (protons and carbon ions). All the boundaries are set to the absorbing boundary condition for particles and fields.

**Acknowledgements**

This work was supported by the Institute for Basic Science under IBS-R012-D1.

**References**

1. Esirkepov, T. *et al.* Highly efficient relativistic-ion generation in the laser-piston regime. Phys. Rev. Lett. **92**, 175003 (2004).

2. Macchi, A., Cattani, F., Liseykina, T. V. & Cornolti, F. Laser acceleration of ion bunches at the front surface of overdense plasmas. Phys. Rev. Lett. **94**, 165003 (2005).

3. Robinson, A. P. L. *et al.* Radiation pressure acceleration of thin foils with circularly polarized laser pulses. New J. Phys. **10**, 013021 (2008).

4. Hegelich, B. M., *et al.* Laser acceleration of quasi-monoenergetic MeV ion beams. Nature (London) **439**, 441-444 (2006).

5. Schwoerer, H. *et al.* Laser-plasma acceleration of quasi-monoenergetic protons from microstructured targets. Nature (London) **439**, 445-448 (2006).

6. Daido, H., Nishiuchi, M & Pirozhkov, A. S. Review of laser-driven ion sources and their applications. Rep. Prog. Phys. **75**, 056401 (2012).

7. Macchi, A., Borghesi, M. & Passoni, M. Ion acceleration by superintense laser-plasma interaction. Rev. Mod. Phys. **85**, 751-793 (2013).

8. Malka, V. *et al.* Principles and applications of compact laser–plasma accelerators. Nat. Phys. **4**, 447-453 (2008).

9. Bulanov, S. V. & Khoroshkov, V. S. Feasibility of using laser ion accelerators in proton therapy. Plasma Phys. Rep. **28**, 453-456 (2002).

10. Tajima, T., D. Habs, D. & Yan, X. Laser acceleration of ions for radiation therapy. Rev. Accel. Sci. Tech. 0**2**, 201-228 (2009).

11. Ledingham, K. W. D., McKenna, P. & Singhal, R. P. Applications for nuclear phenomena generated by ultra-intense lasers. Science **300**, 1107-1111 (2003).

12. Kim, I. J. *et al.* Transition of proton energy scaling using ultrathin target irradiated by linearly polarized femtosecond laser pulses. Phys. Rev. Lett. **111**, 165003 (2013).

13. Snavely, R. A. *et al.* Intense high-energy proton beams from petawatt-laser irradiation of solids. Phys. Rev. Lett. **85**, 2945-2948 (2000).

14. Wilks S. C. *et al.* Energetic proton generation in ultra-intense laser–solid interactions. Phys. Plasmas **8**, 542-549 (2001).

15. Borghesi, M. *et al.* Electric field detection in laser-plasma interaction experiments via the proton imaging technique. Phys. Plasmas **9**, 2214-2218 (2002).

16. Kraft, S. D *et al.* Dose-dependent biological damage of tumour cells by laser-accelerated proton beams. New J. Phys. **12**, 085003 (2010).

17. Fuchs, J. *et al.* Laser-driven proton scaling laws and new paths towards energy increase. Nat. Phys. **2**, 48-54 (2006).

18. Robson, L. *et al.* Scaling of proton acceleration driven by petawatt-laser-plasma interactions. Nat. Phys. **3**, 58-62 (2007).

19. Macchi, A., Veghini, S., Liseykina, T. V. & Pegoraro, F. Radiation pressure acceleration of ultrathin foils. New J. Phys. **12**, 045013 (2010).

20. Pegoraro, F. & Bulanov, S. V. Photon bubbles and ion acceleration in a plasma dominated by the radiation pressure of an electromagnetic pulse. Phys. Rev. Lett. **99**, 065002 (2007).

21. Yu, T-P., Pukhov, A., Shvets, G. & Chen, M. Stable laser-driven proton beam acceleration from a two-ion-species ultrathin targets. Phys. Rev. Lett. **105**, 065002 (2010).

22. Henig, A. *et al.* Radiation-pressure acceleration of ion beams driven by circularly polarized laser pulses. Phys. Rev. Lett. **103**, 245003 (2009).

23. Kar, S. *et al.* Ion acceleration in multispecies targets driven by intense laser radiation pressure. Phys. Rev. Lett. **109**, 185006 (2012).

24. Qiao, B. *et al.* Stable GeV ion-beam acceleration from thin foils by circularly polarized laser pulses. Phys. Rev. Lett. **102**, 145002 (2009).

25. Yan, X. Q. *et al.* Generating high-current monoenergetic proton beams by a circularly polarized laser pulse in the phase-stable acceleration regime. Phys. Rev. Lett. **100**, 135003 (2008).

26. Lee, C.-L., Yang, X. & Greenham, N. C. Determination of the triplet excited-state absorption cross section in a polyfluorene by energy transfer from a phosphorescent metal complex. Phys. Rev. B **76**, 245201 (2007).

27. Choi, I. W. *et al.* Simultaneous generation of ions and high-order harmonics from thin conjugated polymer foil irradiated with ultrahigh contrast laser. Appl. Phys. Lett. **99**, 181501 (2011).

28. Freeman, C. G. *et al.*, Calibration of a Thomson parabola ion spectrometer and Fujifilm plate detectors for protons, deuterons, and alpha particles. Rev, Sci. Instrum. 82, 073301 (2011).

29. Prasad, R. *et al.* Calibration of Thomson parabola-MCP assembly for multi-MeV ion spectroscopy. Nucl. Instrum. Methods Phys. Res., Sect. A **623**, 712-715 (2010).

30. Pae, K. H., Choi, I. W. & Lee, J. Self-mode-transition from laser wakefield accelerator to plasma wakefield accelerator of laser-driven plasma-based electron acceleration. Phys. Plasmas **17**, 123104 (2010).

31. Kim, I J. *et al.*, Relativistic frequency upshift to the extreme ultraviolet regime using self-induced oscillatory flying mirrors. Nat. Commun. **3**, 1231 (2012).

**Figure legends**

Figure 1. **Physical configuration of RPA shown in a 3D-PIC simulation.** (a) Three-dimensional density maps of electrons, protons, and laser radiation pressure at the peak of a 30-fs laser pulse. A circularly polarized laser pulse with an intensity of $6.0 \times 10^{20}$ W/cm$^2$ was applied to a 15-nm-thick target. (b) Total charge density (black) and proton charge density (red), and longitudinal electric field (blue) along the laser propagation axis. These values are the average values over the central region with a 0.8-μm diameter along the laser propagation axis. The total charge density is contributed from electrons, protons, and C$^{6+}$ ions. The target is placed initially at $z = 0$.

Figure 2. **Energy spectra of the protons and carbon ions obtained from ultrathin targets.** (a) Raw image of proton and $C^{6+}$ energy spectra measured with a Thomson parabola spectrometer. (b) Calibrated energy spectra of protons obtained from a 15-nm F8BT target irradiated with a circularly polarized laser pulse of intensity 6.1 × $10^{20}$ W/cm$^2$ ($a_0$ = 17) (red) and from a 20-nm target with a linearly polarized laser pulse of intensity 7.0 × $10^{20}$ W/cm$^2$ ($a_0$ = 18) (black). The background level was estimated along the blue curve located just below the proton trace. The laser contrast was 3 × $10^{-11}$ at 6 ps before the main pulse.

Figure 3. **Optimization of target thickness and scaling to laser intensity for maximum proton energy.** (a) Dependence of maximum proton energy on target thickness. Sold triangles (red) and sold circles (black) denote proton energies obtained from experiments using circularly and linearly polarized laser pulses, respectively, while empty circles (blue) denotes those obtained from 3D-PIC simulations. (b) Dependence of maximum proton energy on laser intensity for circular (red) and linear (black) polarizations. Each point denotes the average value of three laser shots, and the error bar represents the standard deviation.

Figure 4. **RPA features obtained from 3D-PIC Simulations.** (a) Phase space plot of protons within a radius λ around the laser pulse axis at 20 fs (black), 30 fs (red), and 40 fs (blue) after the start of the interaction. The inset shows the temporal profile of a sin$^2$ laser pulse with a duration of 30 fs FWHM. (b) Temporal evolution of maximum proton energy. Red curve represents the case obtained with a 15-nm target accelerated by a circularly polarized laser pulse with an intensity of 6.2×$10^{20}$ W/cm$^2$.

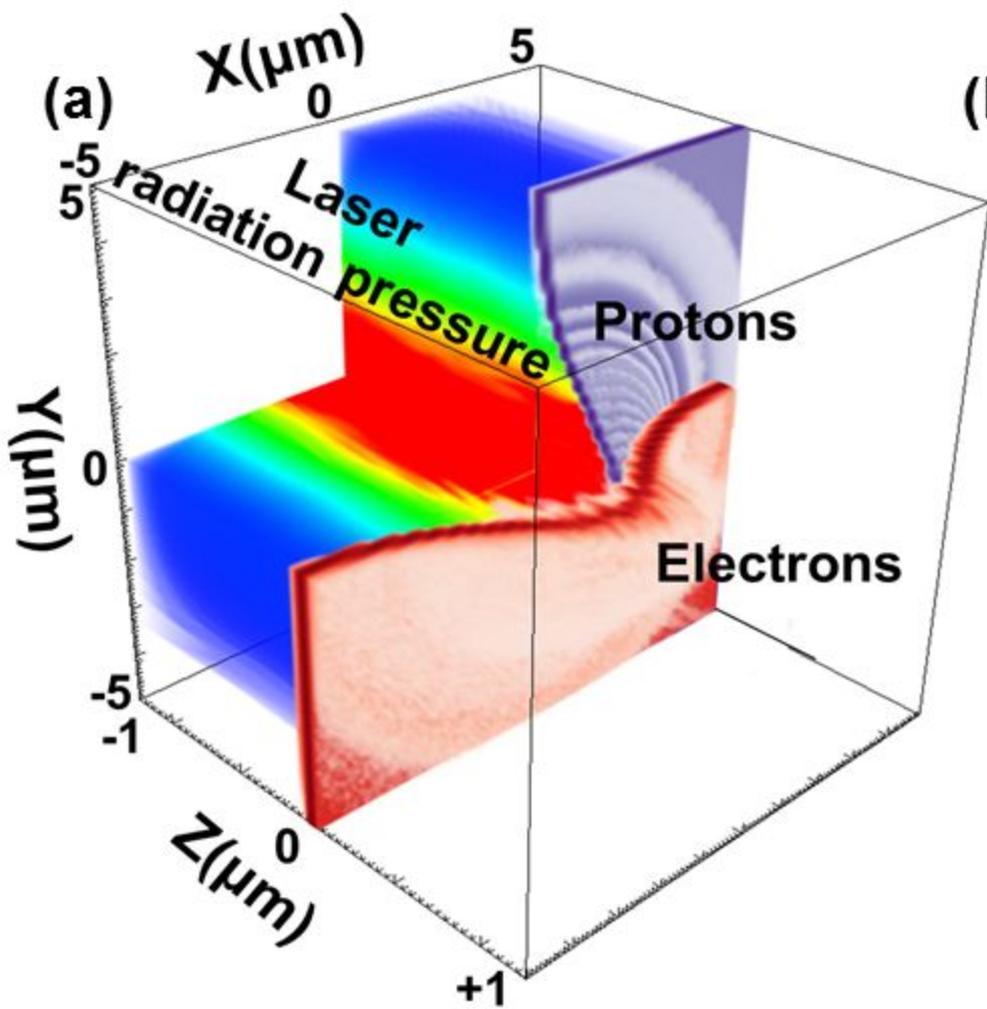 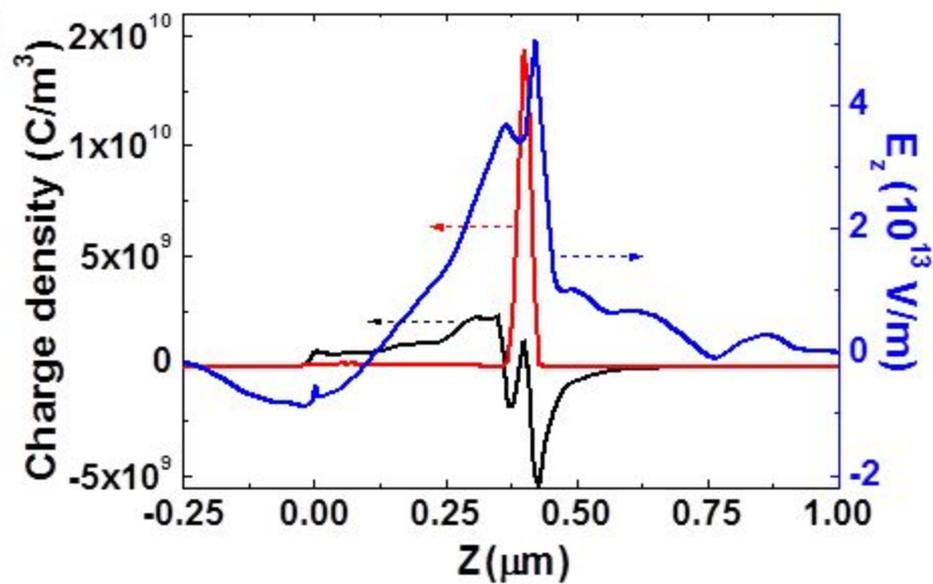

**Figure 1 of 4.**

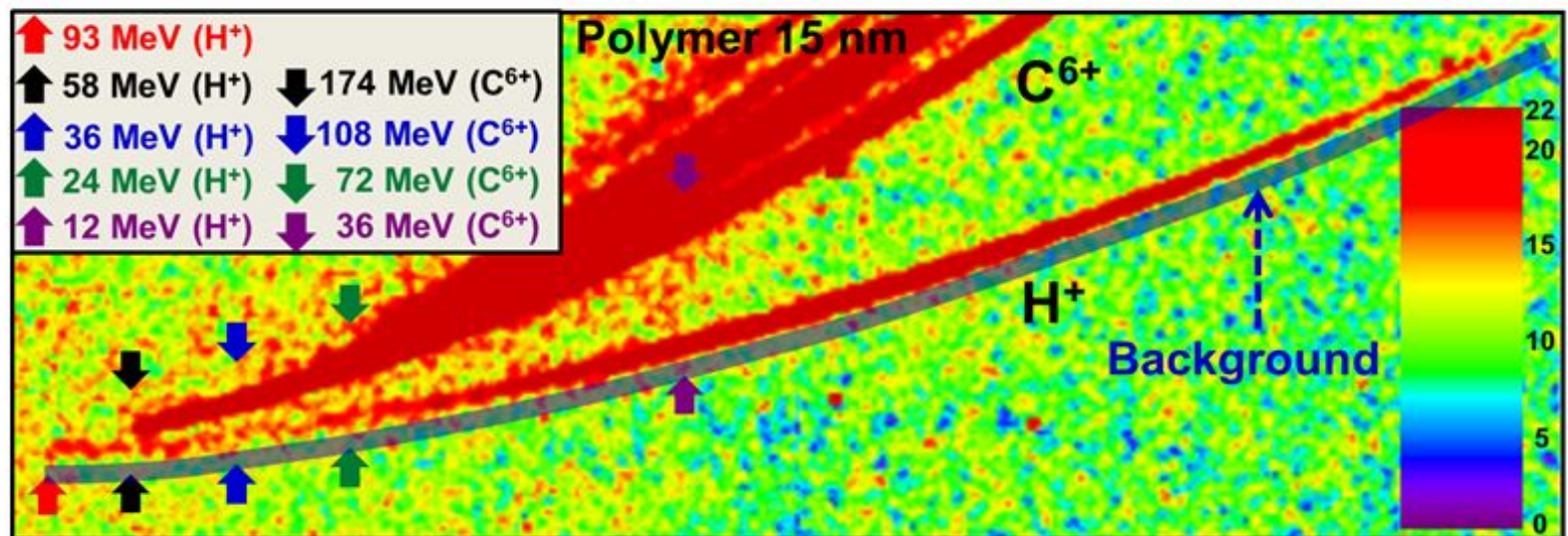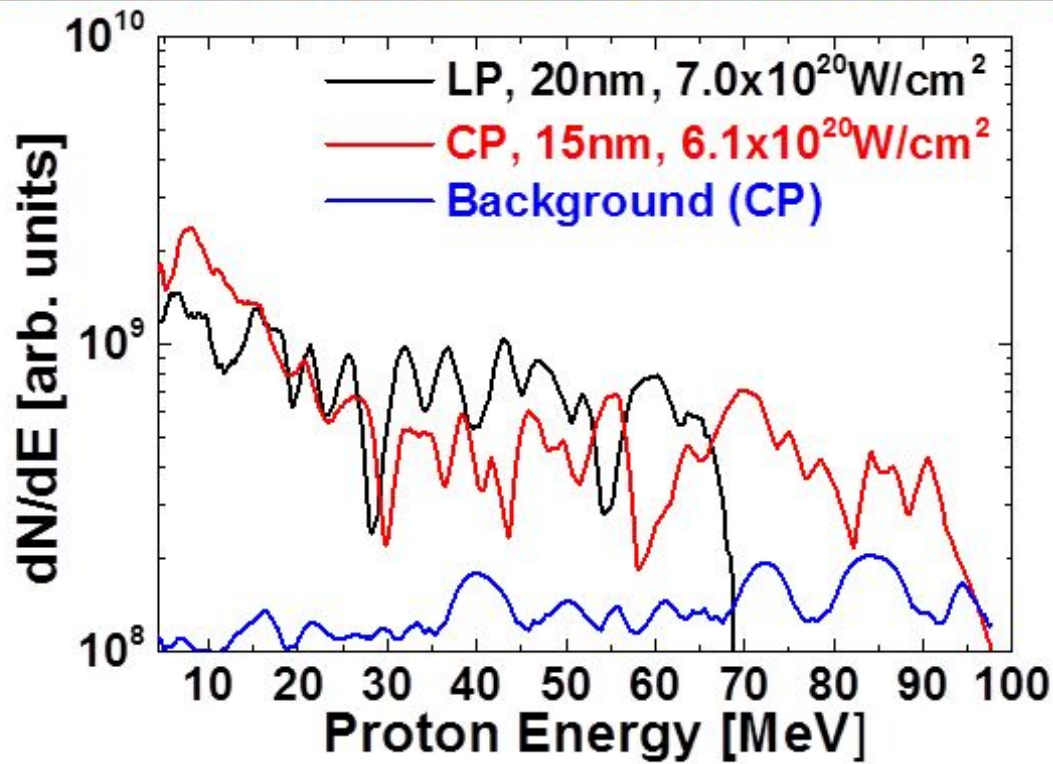

Figure 2 of 4.

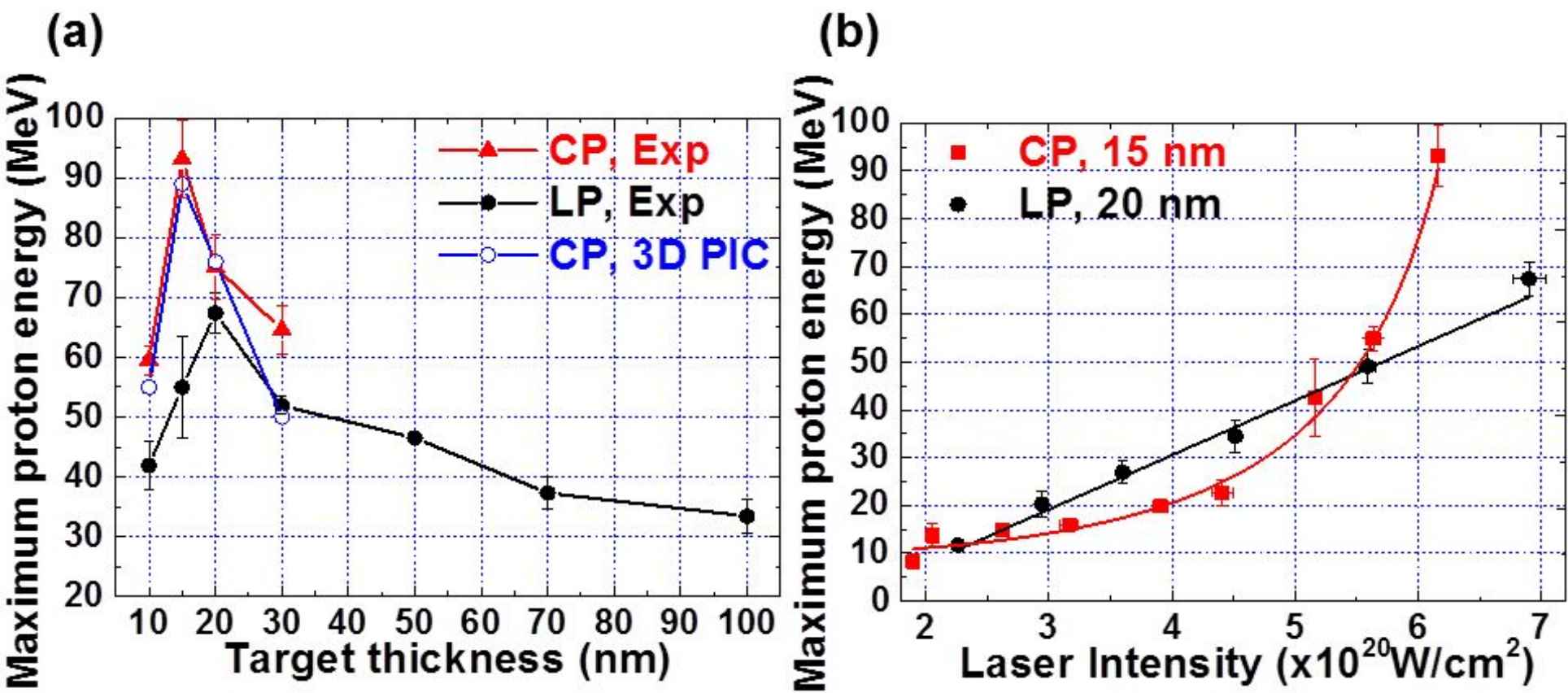

Figure 3 of 4.

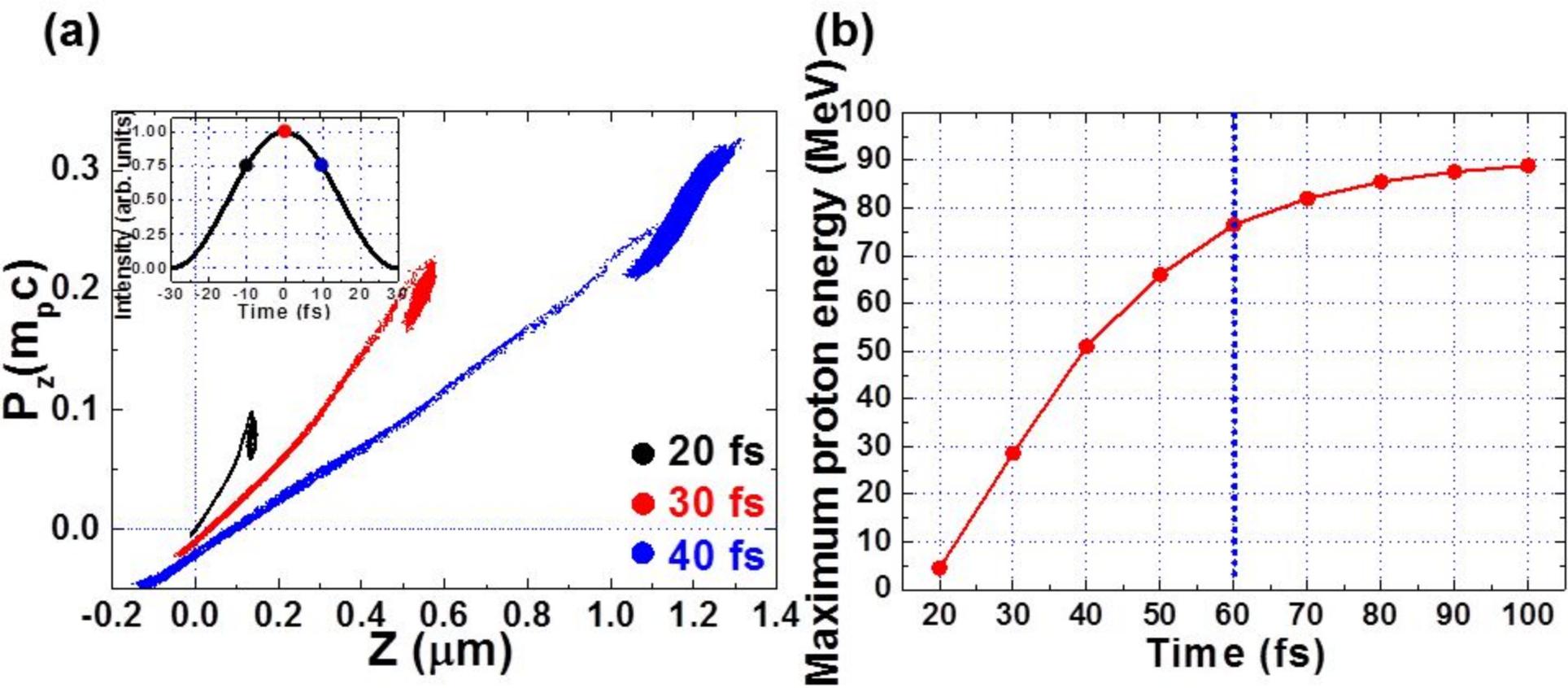

Figure 4 of 4.